\documentclass[aps,prb,amsmath,twocolumn,amssymb,floatfix,showpacs,
superscriptaddress,footinbib]{revtex4-1}
\pdfoutput=1
\usepackage[dvips]{graphics}
\usepackage{bm}
\usepackage{epsfig}
\usepackage{enumerate}
\usepackage{subfigure}
\usepackage{color}
\usepackage{amsmath}
\usepackage{graphicx}
\usepackage{hyperref}
\hypersetup{
    pdfnewwindow=true,      
    colorlinks=true,       
    linkcolor=blue,          
    citecolor=magenta,        
    filecolor=blue,      
    urlcolor=blue        
}

\newcommand{\etal}{{\it et al.~}}
\newcommand{\ie}{{\it i.e., }}

\newcommand{\eps}{{\epsilon }}
\newcommand{\bea}{\begin{eqnarray}}
\newcommand{\eea}{\end{eqnarray}}
\newcommand{\beq}{\begin{equation}}  
\newcommand{\eeq}{\end{equation}}

\begin{document} 
\title{Probing decoupled edge states in a zigzag phosphorene nanoribbon via RKKY exchange interaction} 

\author{SK Firoz Islam}
\email{firoz@iopb.res.in}
\affiliation{Institute of Physics, Sachivalaya Marg, Bhubaneswar-751005, India}
\author{Paramita Dutta}
\email{paramitad@iopb.res.in}
\affiliation{Institute of Physics, Sachivalaya Marg, Bhubaneswar-751005, India}
\author{A. M. Jayannavar}
\email{jayan@iopb.res.in}
\affiliation{Institute of Physics, Sachivalaya Marg, Bhubaneswar-751005, India}
\affiliation{Homi Bhabha National Institute, Training School Complex, Anushakti Nagar, Mumbai 400085, India}
\author{Arijit Saha}
\email{arijit@iopb.res.in}
\affiliation{Institute of Physics, Sachivalaya Marg, Bhubaneswar-751005, India}
\affiliation{Homi Bhabha National Institute, Training School Complex, Anushakti Nagar, Mumbai 400085, India}

\begin{abstract}
Phosphorene is an anisotropic puckered two-dimensional ($2$D) hexagonal lattice of phosphorus atoms. 
The edge modes in a zigzag phosphorene nanoribbon (ZPNR) are quasi-flat in nature and fully isolated from the bulk states,
which are unique in comparison to the other hexagonal lattices like graphene, silicene etc. We theoretically investigate the 
Ruderman-Kittel-Kasuya-Yosida (RKKY) exchange interaction between two magnetic impurities placed on the nanoribbon, 
and extract the signatures of the flat edge states via the behavior of it. Due to the complete separation of the edge states from 
the bulk, we can isolate the edge mode contribution via the RKKY interaction from that of the bulk by tuning the 
external gate potential when both the impurities are placed at the same edge. The bulk induced RKKY interaction 
exhibits very smooth oscillation with the distance between the two impurities, whereas for edge modes it fluctuates 
very rapidly. We also explore the effect of tensile strain both in absence and presence of gate voltage and 
reveal that the RKKY interaction strength can be boosted under suitable doping, when both the impurities are within the bulk. 
\end{abstract}

\maketitle
\section{Introduction}

In recent times, phosphorene has emerged as a promising $2$D material in regards of its potential applications
in nanoelectronics owing to the unusual anisotropic band structure~\cite{liu2014phosphorene,castellanos2016all,koenig2014electric}. 
It is a bilayer puckered hexagonal lattice of black phosphorus exhibiting both the linear and quadratic energy dispersion in the bulk, 
depending on the direction of the quasi-particle's momentum. This anisotropy in band structure has recently been exploited
in a series of theoretical works, especially in the context of transport
properties~\cite{PhysRevB.96.035422,wang2015highly,PhysRevB.95.144515,PhysRevB.95.045422}. 
Apart from the bulk, zigzag phosphorene nanoribbon (ZPNR) can possess two quasi-flat edge modes which are
completely isolated from the conduction and valence 
bands~\cite{PhysRevB.94.075422,ezawa2014topological,PhysRevB.91.085409,PhysRevB.94.125410,PhysRevB.93.075408}. This is 
in complete contrast to the case of other existing $2$D hexagonal lattice
structures~\cite{RevModPhys.81.109,PhysRevB.92.035413,PhysRevB.96.045418} where the edge modes merge into the bulk at the two Dirac points. 
The origin of such decoupled nature of the flat bands in ZPNR is due to the presence of two out-of plane zigzag chains, coupled by relatively strong hopping parameter, 
which has recently been addressed by Ezawa~\cite{ezawa2014topological}. It has also been pointed out that two edge modes can be separated from each other by applying 
a suitable gate voltage between two opposite transverse edges of the zigzag chain~\cite{ezawa2014topological}. 
The 2D phosphorene materials have several advantageous features over the other existing 2D materials, 
such as phosphorene based field effect transistor (FET) can be a more suitable device in comparison
to a graphene based FET, especially in regards of switching on/off
ratio~\cite{koenig2014electric,das2014ambipolar,li2014black}. Moreover, charge carriers in phosphorene can acquire very high mobility ($\sim1000$ cm$^2$/Vs) in comparison to  
transition metal dichalkogenides materials~\cite{koenig2014electric,li2014black,xia2014rediscovering} ($\sim200$ cm$^2$/Vs) at room temperature. 

The RKKY interaction~\cite{ruderman1954indirect,kasuya1956electrical,yosida1957magnetic} between two magnetic impurities 
is an indirect exchange interaction mediated by the conduction electrons of the host material. This interaction plays the 
key role in determining the magnetic ordering in some electronic systems such as spin glasses~\cite{PhysRevB.51.15250} 
and alloys~\cite{PhysRevB.36.492}. The RKKY interaction has been studied very extensively in various Dirac materials like 
graphene~\cite{PhysRevB.88.085405,PhysRevLett.99.116802,PhysRevB.72.155121,
PhysRevB.84.115119,PhysRevB.77.195428,PhysRevB.81.205416,PhysRevB.83.165425,PhysRevB.84.125416},
bilayer graphene~\cite{PhysRevB.87.165429,PhysRevLett.101.156802}, carbon nanotube\cite{PhysRevB.87.045422,PhysRevB.92.035411},
silicene~\cite{xiao2014ruderman,PhysRevB.94.045443}, topological insulator~\cite{PhysRevLett.106.097201,PhysRevB.96.024413} etc. It can be 
probed by several methods like the single-atomic magnetometry of a pair of magnetic atoms~\cite{khajetoorians2012atom,zhou2010strength} 
and magnetotransport measurement based on angle-resolved photo-emission spectroscopy (ARPES)~\cite{PhysRevLett.91.116601}. 
Apart from these, a method of directly probing the local spin susceptibility, compatible with 1D nanoribbon, has also been 
proposed in Ref.~[\onlinecite{PhysRevB.88.045441}]. Very recently, the features of RKKY interaction has been proposed to probe the electrically controlled 
zero energy conducting edge mode in the topological phase of buckled hexagonal silicene lattice structure~\cite{PhysRevB.94.045443}.

Till date, several anisotropic electronic transport properties of phosphorene, as mentioned earlier, have been reported.
Nevertheless, the magnetic exchange interaction in presence of magnetic impurities is still under consideration of 
theoretical investigation as far as phosphorene is concerned. In very recent works, the RKKY exchange interaction
has been considered in the bulk of phosphorene, aiming to explore the effect of anisotropy of the band
dispersion~\cite{NJP_phos,reja}. However, the signatures of unusual quasi-flat edge modes in ZPNR have not been
explored so far in the context of RKKY interaction, althogh room temperature magnetism has been explored
in details in Ref.~[\onlinecite{PhysRevB.94.075106}]. Apart from the anisotropic nature of the interaction
in bulk phosphorene~\cite{NJP_phos,reja}, the edge modes may play a vital role in the RKKY interaction in ZPNR.
Motivated by this, in this article, we investigate the behavior of RKKY exchange interaction in ZPNR and
extract the responses of quasi-flat edge modes from it.

In our work, we consider
two magnetic impurities which are placed either at the same zigzag edge or in the interior of a ZPNR. The features of the quasi-flat edge modes 
in the RKKY interaction are extracted from our numerical results based on the real space Green's function of the system. We observe that the
RKKY interaction between two magnetic impurities placed at the same edge of an undoped nanoribbon is much stronger in comparison
to the case when any one or both of the impurities are away from the edge. Similar to the other $2$D materials, the nature of the interaction 
is oscillatory with the distance between the two impurities. Moreover, a gate voltage applied between two nearest zigzag chains, lying at different planes, 
provides us another degree of freedom to tune the edge modes~\cite{PhysRevB.94.125410} and subsequently RKKY interaction in ZPNR. 
We show that the strength of the exchange interaction can be significantly enhanced by tuning the gate voltage in undoped ZPNR. It depends on the locations
of the impurities as well. 

On the other hand, application of strain has significant influences on the band structure
as well as topological properties of phosphorene. Very recently, it has been predicted that the application of a tensile 
or in-plane strain in spin-orbit coupled phosphorene can close and reopen the band gap and gives rise to the topological phase transition~\cite{peeters_strain}. 
Motivated by this prediction, we also examine the effect of strain on the RKKY interaction both in absence and presence of the gate voltage. 
However, we do not consider spin-orbit coupling in our ZPNR as so far there is no experimental evidence of spin-orbit interaction in monolayer phosphorene.
Moreover, we are interested in probing the detached edge modes rather than topological features.
The application of a tensile strain can induce a curvature to the band structure for which the RKKY interaction acquires a phase. For all the three possible configurations 
of the location of the impurities {\it i.e.}, both are at the edge or away from the edge or one at the edge considering the other one within the bulk of the ribbon, we present 
our results of RKKY interaction to understand the effect of strain. Interestingly, under suitable doping condition, the exchange interaction can be affected by tuning 
the degree of strain. 
On the contrary, the combined effect of the gate voltage and strain on the RKKY interaction yields non-significant conribution when both the impurities are 
situated within the interior of the nanoribbon.

The remainder of the paper is organized as follows. In Sec.~\ref{sec2}, we introduce the lattice structure and the tight binding 
Hamiltonian for phosphorene with the inclusion of gate voltage and strain. 
Sec.~\ref{sec3} is devoted to the analysis of band structure of ZPNR under the influence of the gate voltage and strain. 
A brief discussion on the Green function formalism for analysing the RKKY interaction is given in Sec.~\ref{sec4}. 
Our numerical results of the RKKY interaction as a function of the distance between the two magnetic impurities, both in absence and presence 
of the gate voltages and the tensile strain, are presented in Sec.~\ref{sec5}. Finally, we summarize and conclude in Sec.~\ref{sec6}.

\section{Tight binding Model Hamiltonian}\label{sec2}
In this section, we first provide a short description of the lattice geometry of phosphorene. The puckered hexagonal lattice 
of phosphorene is very similar to that of graphene but with two nearest neighbor zigzag chains lying at two different parallel planes. 
Unlike graphene, the bond lengths as well as corresponding hopping parameters are not identical to each other. 
It depends on the plane as well as the sublattice of the ribbon.
\begin{figure}[!thpb]
\centering
\includegraphics[height=6.2cm,width=0.80 \linewidth]{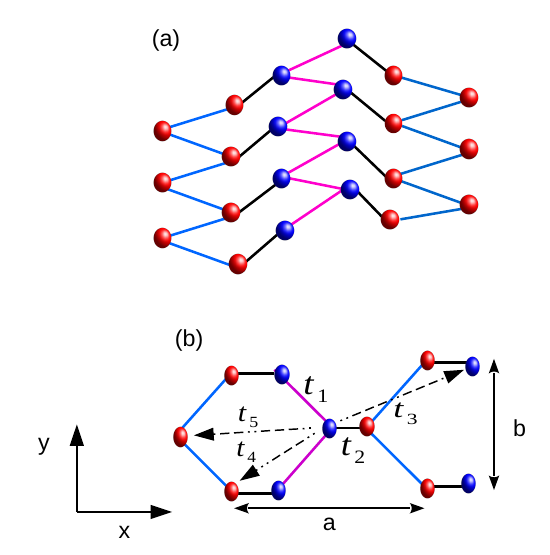}
\caption{(Color online) (a) Schematic view of the phosphorene lattice structure is presented.
Two different colors, blue and red, are used to denote the atoms belonging to two different planes.
(b) A projected view of the lattice in $x$-$y$ plane is demonstrated. 
The five non-equivalent hopping parameters associated with the lattice are marked by $t_{1}$, $t_{2}$, $t_{3}$, $t_{4}$ and $t_{5}$. 
$a$ and $b$ are the two lattice constants along the $x$ and $y$ directions respectively.}
\label{lattice}
\end{figure}
A typical sketch of phosphorene lattice structure is depicted in Fig.~\ref{lattice}(a). Also, a top view projected in the $x$-$y$ plane 
is shown in Fig.~\ref{lattice}(b). Corresponding to the position vector of $i$-th atom $r_i$, we denote the hopping parameter by $t_i$. The 
different structure parameters associated with this lattice structure can be found
in Refs.~[\onlinecite{peeters_strain,qiao2014high}]. 
The lattice parameters are given as $r_1=22.40$ nm, $r_2=22.80$ nm; ($r_{1x},r_{1y},r_{1z}$)=($15.03,16.60,0$) nm and 
($r_{2x},r_{2y},r_{2z}$)=($7.86,0,21.40$) nm. Other coordinates $r_3,r_4,r_5$ can simply be obtained from $r_1$ and $r_2$. 
The two lattice constants in $x$-$y$ plane are $a=45.80$ nm and $b=33.20$ nm.
 
The tight binding Hamiltonian of this puckered lattice, as proposed in Ref.~[\onlinecite{katsnelson}], in absence of spin-orbit 
interaction is given by
\begin{equation}
 H_0=\sum_{ij}t_{ij}c_{i}^{\dagger}c_j \ ,
\label{H}
\end{equation}
where the summation in Eq.(\ref{H}) runs upto the fifth nearest neighbor and $t_{ij}$ is the hopping parameter between
$i$-th and $j$-th atom. The creation (annihilation) operators at $i$-th cite are denoted by $c_{i}^{\dagger}$ ($c_{i}$). 
The numerical values of the hopping parameters are~\cite{katsnelson,peeters_strain}: $t_1=-1.22$ eV, $t_2= 3.665$ eV, $t_3=-0.205$ eV, 
$t_4=-0.105$ eV and $t_5=-0.055$ eV. 

\subsection{Inclusion of gate voltage}
As the system is composed of two parallel planes of zigzag chain, an application of suitable gate voltage 
between two opposite edges but in different planes can modify the band structure as pointed out by Ezawa~\cite{ezawa2014topological}
and Ma \etal~\cite{PhysRevB.94.125410} in ZPNR. Note that, in-plane hopping parameters are all negative while
the inter-plane hopping parameter ($t_2$) is positive. However, in order to tune the full band dispersion
with respect to the Fermi level, one can bias the top and bottom planes as $U_t=U$ and
$U_b=-U$ respectively. The latter gives rise to an additional band gap $\Delta_g=2U$.
This kind of bias can be realized experimentally~\cite{zhang2009direct}. 

Now, including the effect of the gate voltage, the total Hamiltonian of the system can be written as 
\begin{eqnarray}\label{hamil_tight}
 H&=&\sum_{ij}t_{ij}c_{i}^{\dagger}c_j+\sum_{i}Uc_{i}^{\dagger}c_{i}\ .
\end{eqnarray}

Here, in our analysis, we bias only the top plane by $U$ and consider the bottom plane at $U=0$. In the second term of Eq.(\ref{hamil_tight}), 
the index `$i$' runs over all the sublattices of the zigzag chain in the top plane only.

\subsection{Inclusion of strain}
The strain has a very significant impact on the band structure of 2D sheet of phosphorene. As mentioned previously,
phosphorene with spin-orbit coupling can undergo from normal to topological insulator phase transition under suitable in-plane
or perpendicular tensile strain~\cite{peeters_strain}. However, in our case even without spin-orbit coupling, the strain modulates 
the band structure by modifying the hopping parameters and hence a significant influence on the RKKY magnetic exchange interaction is expected.

When strain is applied, the initial geometrical parameters are deformed as $(r_{ix},r_{iy},r_{iz})=[(1+\eps_x)r_{ix}^{0}
,(1+\eps_y)r_{iy}^{0},(1+\eps_z)r_{iz}^{0}]$, where $\eps_j$ is the strain along $j$-th direction.
In the linear deformation regime, $r_i$ can be simplified up to the first order as
\begin{equation}
 r_i=(1+\kappa_x^i\eps_x+\kappa_{y}^{i}\eps_y+\kappa_z^i\eps_z)r_i^0\ ,
\end{equation}
with $\kappa_j^i=(r_{ij}/r_{i}^{0})^2$ being the coefficients related to the structural parameters of phosphorene.
Finally, following Harrison relation~\cite{harrison}, one can obtain the strain induced modified hopping parameters as
\begin{equation}
 t_{i}\simeq(1-2\kappa_x^i\eps_x-2\kappa_y^i\eps_y-2\kappa_z^i\eps_z)t_i^{0}\ .
\end{equation}
However, as it has already been pointed out that the band structure is more sensitive to the perpendicular strain rather
than in-plane strain~\cite{peeters_strain}, in our analysis we only consider the case $\eps_z\neq 0$, while $\eps_x=\eps_y=0$.

\section{Band dispersion}\label{sec3}
In order to find the energy band dispersion of ZPNR (finite along $x$ and infinite along $y$-direction),
\begin{figure}[!htb]
\centering
\includegraphics[width=.5\textwidth,height=5cm]{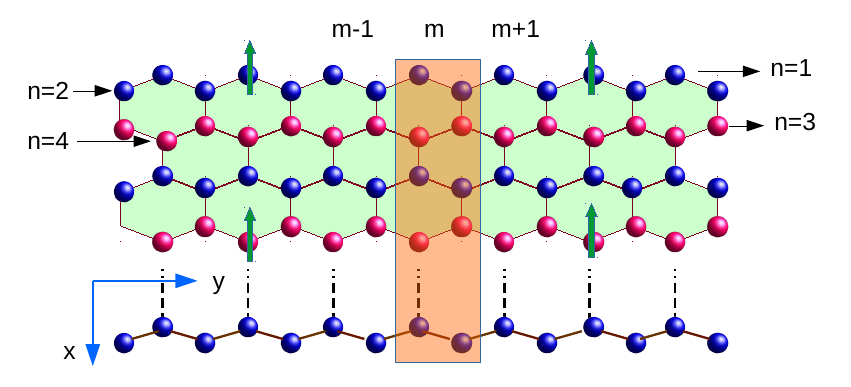}
\caption{Schematic of ZPNR with the atoms of two different zigzag chains denoted by two different colors, blue (dark gray) and pink (light gray).
The magnetic impurities are denoted by vertical green (light gray) arrow sign. 
The line numbers along the $x$ and $y$-direction are denoted by the index $n$ and $m$, respectively.}
\label{ribbon}
\end{figure}
we write an effective difference equation analogous to the case of an infinite one-dimensional chain~\cite{dutta}.
To implement this, the nanoribbon can be considered to consist of an array of the unit cells as shown by the rectangular shaped
orange shadowed region in Fig.~\ref{ribbon}. The width of the zigzag ribbon is determined by the number of atoms $N$ 
per unit cell. The effective difference equation of the ZPNR takes the form as
\bea
(E \mathcal{I}-\mathcal{E}) \psi_m=\mathcal{T} \psi_{m+1}+\mathcal{T}^{\dagger} \psi_{m-1}\ ,
\eea
where
\begin{align}
\psi_m &=\begin{bmatrix} 
\psi_{m,1}\\ \psi_{m,2} \\ \vdots \\ \psi_{m,N} \end{bmatrix}.
\end{align}
\begin{figure}[!thpb]
\centering
\includegraphics[height=6cm,width=\linewidth]{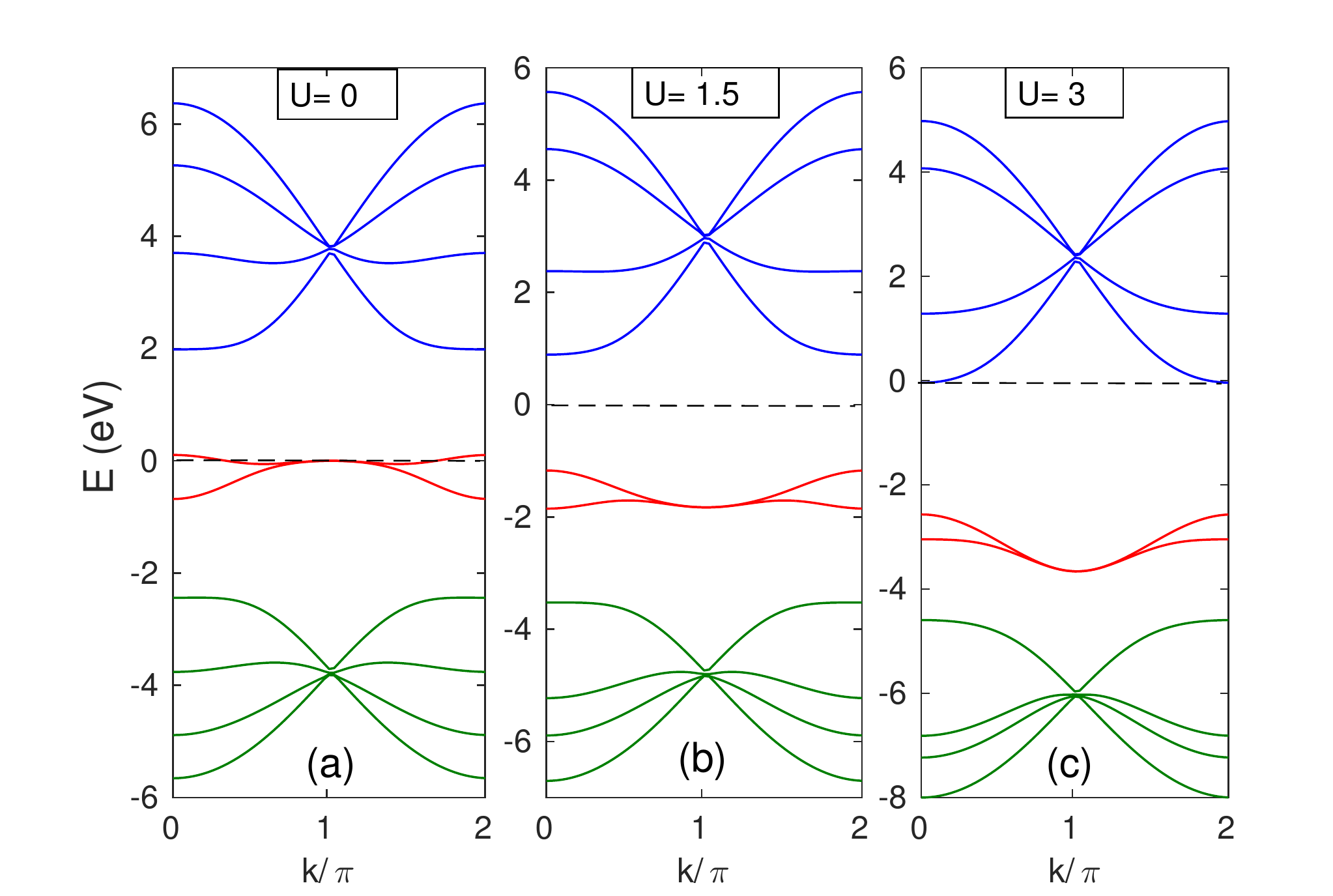}
\caption{(Color online) Energy band dispersion of ZPNR for three different values of gate voltage (in units of $t_1$) is shown. 
The width of the ribbon is taken as $N=10$. The position of the Fermi level is denoted by the dashed line. Here, $\epsilon_{z}=0$.}
\label{band1}
\end{figure}
$\mathcal{E}$ and $\mathcal{T}$ are the on-site energy and nearest-neighbor hopping matrices of the unit cells,
respectively. $\mathcal{I}$ is the identity matrix of dimension $N \times N$. As the zigzag chain is
translationally invariant along $y$-direction, the momentum along that direction ($k$) is conserved and acts as 
a good quantum number. Finally, applying Bloch's theorem the total Hamiltonian of the ZPNR 
can be expressed as
\begin{equation}
(E \mathcal{I}-\mathcal{E})=\mathcal{T} e^{i k b}+\mathcal{T}^{\dagger} e^{-i k b}\ .
\end{equation}
with $b$ as the unit cell separation.
The above equation can be solved numerically to yield energy dispersion of the nanoribbon. 

In Fig.~\ref{band1}, we show the energy band dispersion of a ZPNR of width $N=10$ for three different values of gate voltage 
(a) $U=0$, (b) $U=1.5$, and (c) $U=3$ (in units of $t_1$). A pair of edge modes (red color), decoupled from the bulk band, 
appear in the spectra. This is due to the finite width of the  ZPNR. We observe that the application of gate voltage causes 
the shifting of the whole band (consisting of bulk and edge band) by some finite values of energy being proportional to the external 
gate voltage. Moreover, one of the edge states, which was almost flat in absence of $U$, is deformed to the curved one for $U\neq0$. 
Whereas, in presence of finite $U$, the shape of the other edge state is changed from concave to convex maintaining the degenerate 
or crossing point ($k=\pi$) unchanged.

\begin{figure}[!thpb]
\centering
\includegraphics[height=6cm,width=\linewidth]{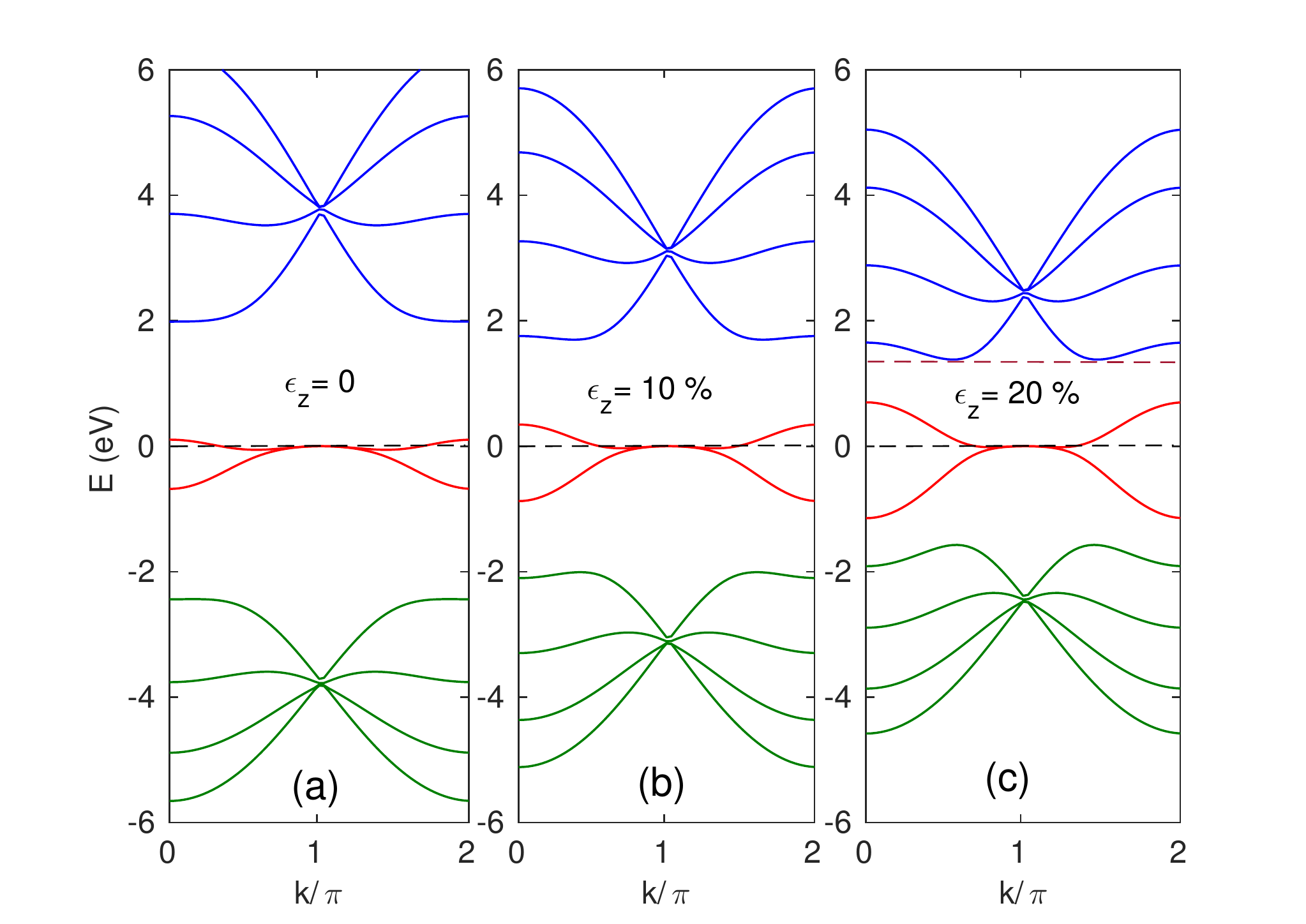}
\caption{(Color online) The features of energy band dispersion of ZPNR for three different strengths of tensile strain is illustrated when $U=0$. 
The width of  the ribbon is considered to be the same as mentioned in Fig.~\ref{band1}. 
Two different positions of the Fermi level is denoted by a dashed line.}
\label{band2}
\end{figure}

In Fig.~\ref{band2}, we demonstrate the energy band dispersion of ZPNR under the influence of perpendicular tensile strain. 
Here, (a), (b), and (c) correspond to different strengths of the strain as $\epsilon_z=0$, $10\%$, and $20\%$, respectively. 
We note that unlike the case of the gate voltage, the tensile strain does not manifest any 
significant shift of the entire band rather it induces a curvature to the bulk modes, leading to the reduction
of the band gap. On the other hand, it widens the gap between two edge modes except at $k=\pi$. 

\begin{figure}[!thpb]
\centering
\includegraphics[height=6cm,width=\linewidth]{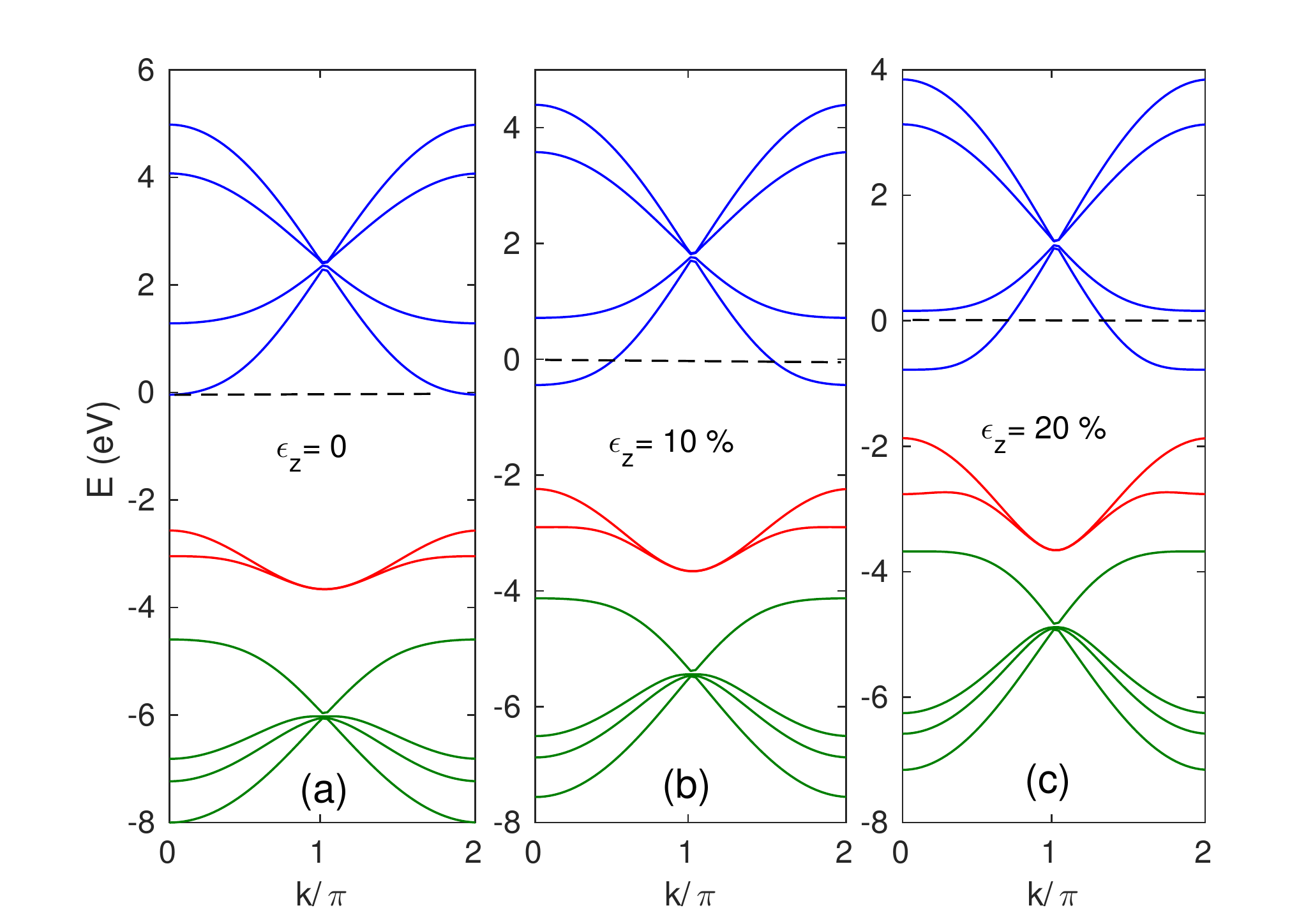}
\caption{(Color online) Energy band dispersion of ZPNR is illustrated for three different strengths of tensile strain in presence of a fixed
non-zero gate voltage $U= 3t_1$. The width of the ribbon is same as mentioned in Fig.~\ref{band1}. The position of the Fermi level is marked by 
the dashed line.
}
\label{band3}
\end{figure}

Finally, we illustrate the band dispersion of ZPNR in presence of both gate voltage and strain in Fig.~\ref{band3}.
Here, we consider the gate voltage to be fixed at $U=3t_1$ and vary the strain as $\epsilon_z=0$, $10\%$, and $20\%$
in (a), (b) and (c) respectively. 
However, in this case the band dispersion appears to be less sensitive to the strain compared to the case in Fig.~\ref{band2}. 
The issue of band gap reduction or band curvature of the bulk states seems to be insensitive to the combined effects of strain and gate voltage.
However, the edge modes still preserve the curvature under the influence of the strain even in presence of the gate voltage. Additionally, the inter 
band separation within the bulk band changes with the enhancement of strain for a finite gate voltage.

\begin{figure*}[!thpb]
\centering
\includegraphics[height=4.5cm,width=0.33\linewidth]{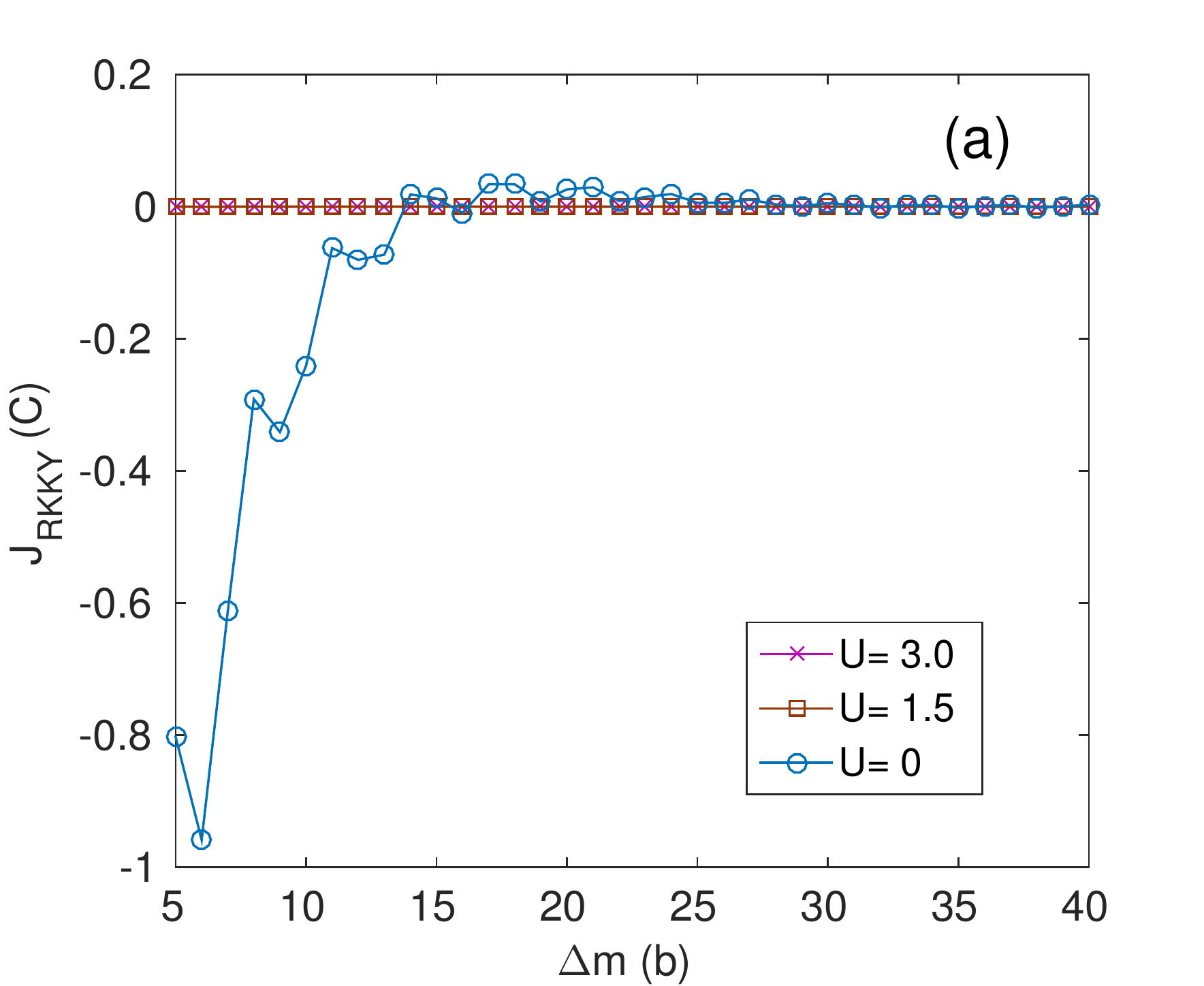}
\includegraphics[height=4.5cm,width=0.32\linewidth]{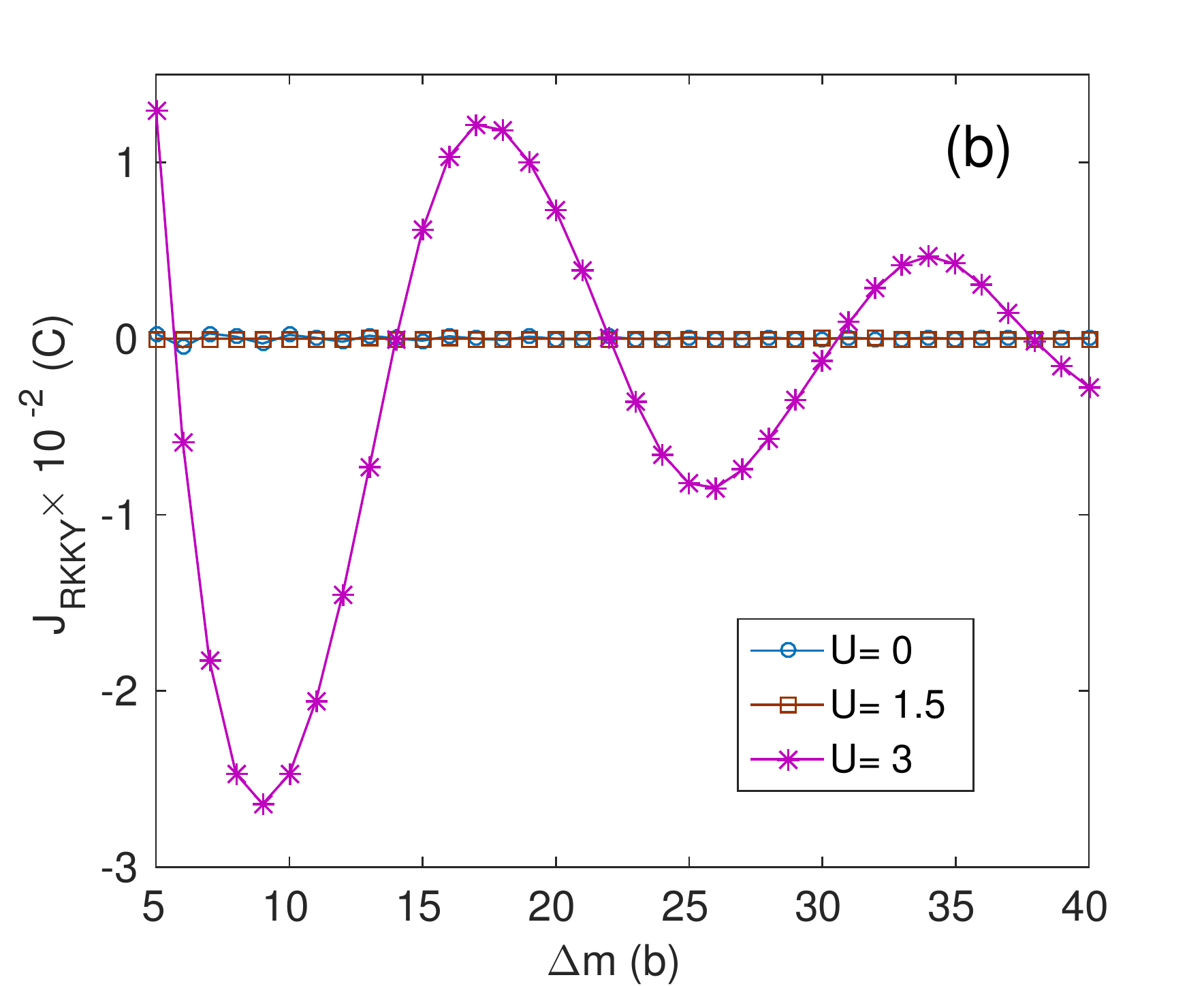}
\includegraphics[height=4.5cm,width=0.33\linewidth]{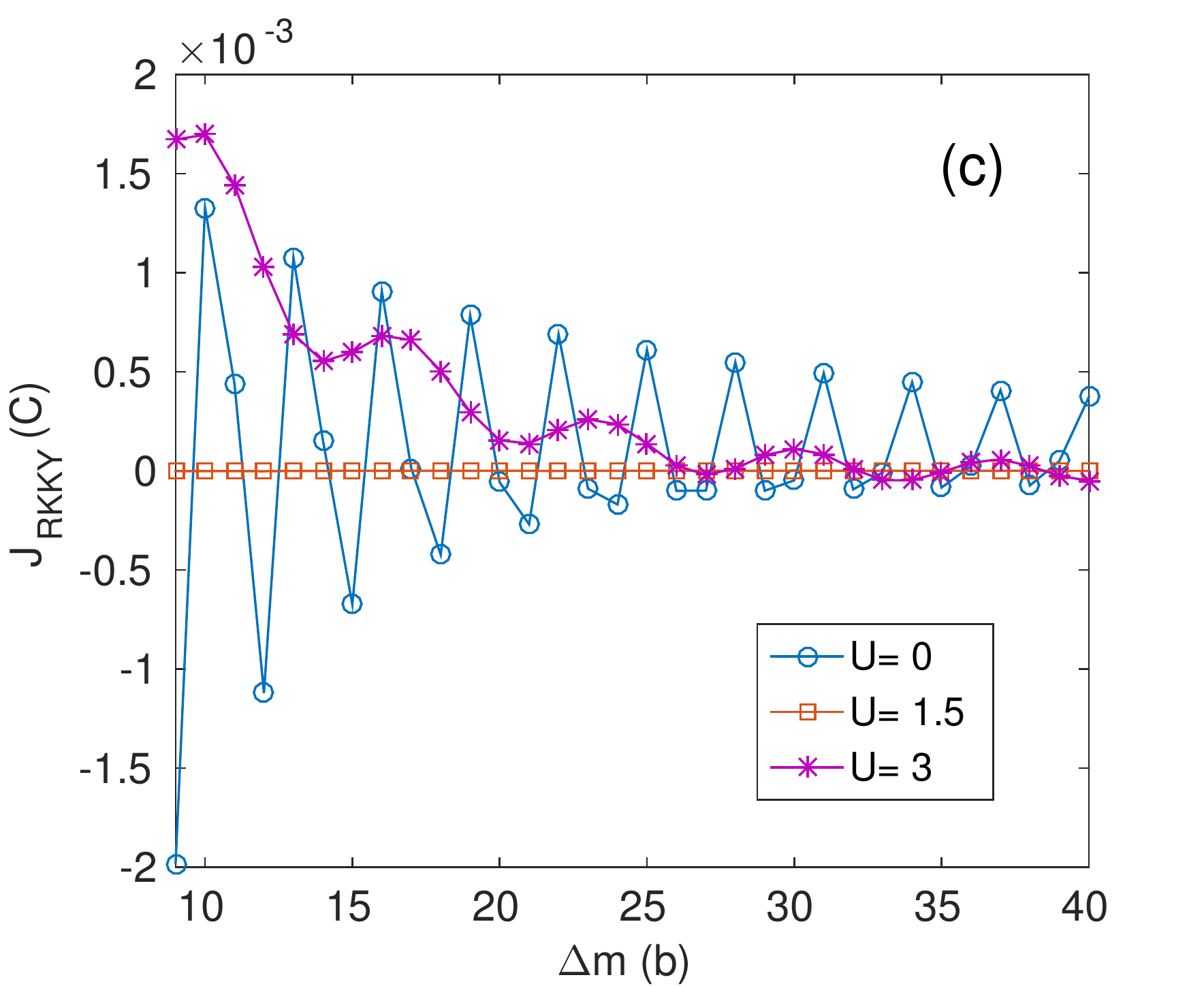}
\caption{(Color online) The behavior of RKKY exchange interaction (in units of $C=(J_c\hbar/2)^2$) between two magnetic
impurities is demonstrated as a function of the distance (in units of the lattice parameter $b$) between them when 
(a) both the impurities are at the same edge, (b) both are within the bulk of ZPNR, and (c) one at the edge, while the other
one is located in the interior of the ZPNR. The size of the undoped ribbon is $10 \times 100$.
For each case, three different gate voltages, $U=0$, $1.5$ and $3$ (in units of $t_1$) are considered.
Here, we choose the strain $\eps_z=0$.}
\label{gate}
\end{figure*}
\section{Theoretical formalism of RKKY interaction}\label{sec4}
In our analysis, we consider the two magnetic impurities located at ($m_1,n_1$) and $(m_2,n_2)$ sites (following the notaions of 
Fig.~\ref{ribbon}) of the nanoribbon. The indirect exchange interaction between these two magnetic impurities is
mediated by the conduction electrons of the host material. The Hamiltonian for the exchange interaction between the spin 
of the magnetic impurity ($S$) and the conduction electron ($s$) can be written as
\beq
 H_{\rm int}= J_{c}\sum_{\alpha}S_{\alpha}.s_{\alpha}\ ,
\eeq
where $\alpha$ is the sublattice index. By implementing the well-known RKKY perturbation theory, the exchange interaction
energy between the spins of two magnetic impurities can be expressed in terms of the Heisenberg form as
~\cite{ruderman1954indirect,kasuya1956electrical,yosida1957magnetic,PhysRevB.81.205416,PhysRevB.94.045443}
\begin{equation}
E({\mathbf{r}})=J_{\alpha\beta}({\mathbf{r}}) S_{\alpha}.S_{\beta}\ .
\end{equation}
Here, one of the two impurities is located at the origin and the other one at position ${\mathbf{r}}$. 
Here, $\alpha$ and $\beta$ represent the sublattice index on which magnetic impurities are placed and $J_{\alpha\beta}$ is the strength 
of the exchange coupling between the two impurities which is linked to the spin-independent susceptibility $\chi_{\alpha\beta}$ as
\beq
J_{\alpha\beta}=C\chi_{\alpha\beta} \ ,
\label{J_equation}
\eeq
where $C=(J_c\hbar/2)^2$. The static susceptibility can be evaluated from the retarded Green's function as
\begin{equation}
\chi_{\alpha \beta}({\mathbf{r,r'}})=-\frac{2}{\pi} {\rm Im} \int_{-\infty}^{E_F}dE [G_{\alpha\beta}^0({\bf {r, r'}},E)
G_{\alpha\beta}^0({\bf{r', r}},E)].
\end{equation}
Here, $G_{\alpha\beta}^{0}$ is the spin-independent unperturbed single particle Green's function, which can
be expressed in the spectral representation as
\begin{equation}
G_{\alpha\beta}^{0}({\bf r, r'},E)=\sum_{n}\frac{\psi_n^{\alpha}(\mathbf{r})\psi_{n}^{\beta}(\mathbf{r'})}{E-E_{n}+i\eta},
\end{equation}
where $n$ runs over all the eigenstates which has to be evaluated by diagonalising Eq.~(\ref{hamil_tight})
for a finite size lattice. 
\section{Numerical results and Discussion}\label{sec5}
In this section, we present our numerical results of the RKKY interaction ($J_{\rm{RKKY}}$), in units of 
$C$, between the two magnetic impurities for various combinations of their locations in ZPNR. We consider 
three different situations when both the impurities are located at the same edge of the nanoribbon or
they are situated within the bulk or one impurity is located at the edge, while the other one is situated
in the interior of the ribbon. We discuss the effect of gate 
voltage, tensile strain and their combination in three different subsections. The size of the ZPNR is considered as: 
the length $M=100$ and width $N=10$. Note that, any further increase of the length of the ZPNR will not 
alter the qualitative nature of our main results. Similarly, the higher value of $N$ does not modulate
the RKKY interaction significantly for the undoped situation. The reason can be attributed to the fact
that, if we enhance the width of the ribbon, the number of bulk modes increases without affecting the
edge states. On the other hand, the RKKY interaction for the undoped condition is strongly dependent
on the behavior of the edge modes. Hence, even for wider ribbon, our results will change quantitatively 
while the qualitative features will remain unaffected.
\begin{figure}[!thpb]
\centering
\includegraphics[height=6.5cm,width=\linewidth]{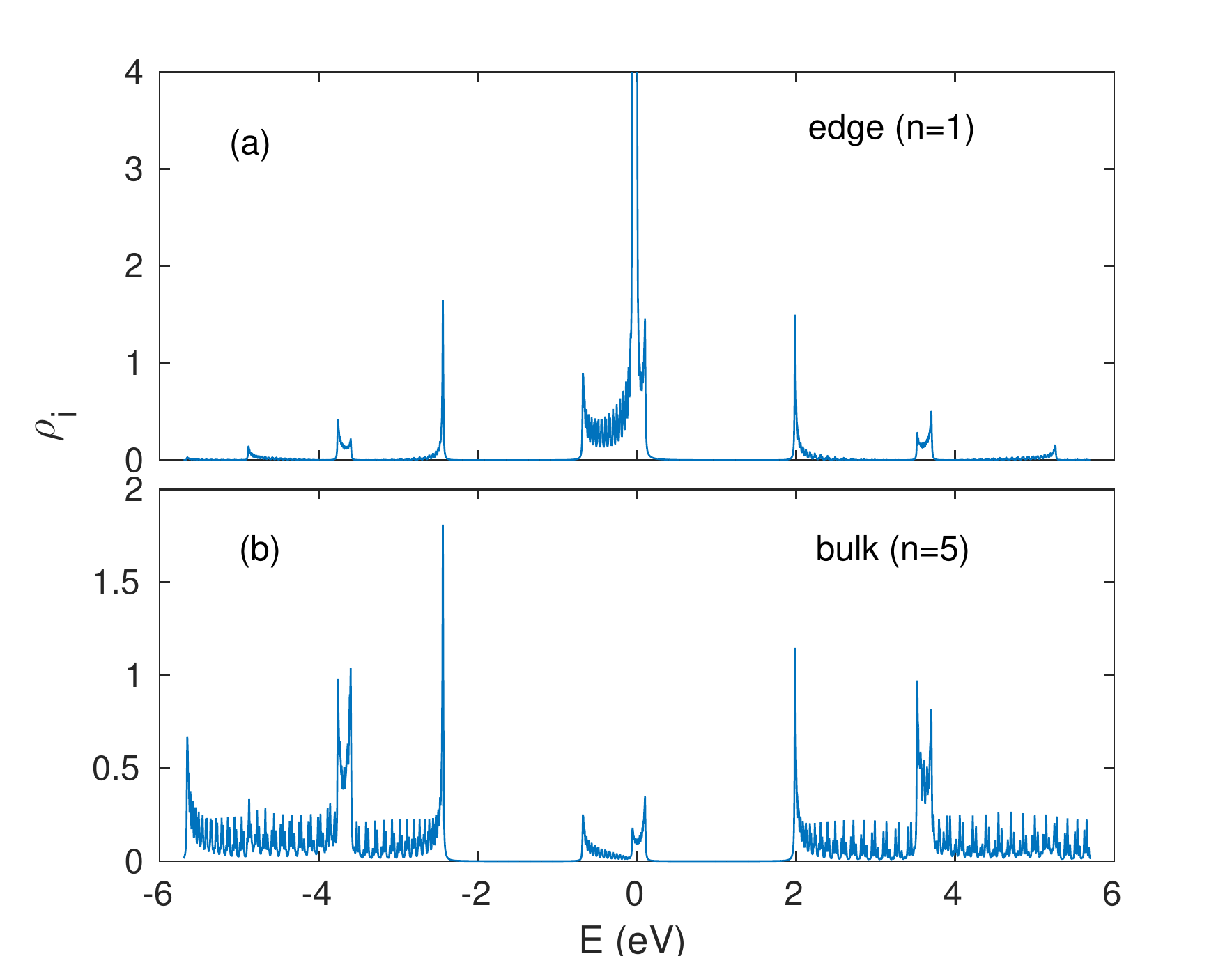}
\caption{(Color online) The behavior of LDOS is shown as a function of energy for (a) edge states ($n=1$) and (b) bulk states ($n=5$).
Here, we have considered $m=26^{th}$ unitcell.}  
\label{ldos}
\end{figure}
\subsection{Effect of external gate voltage}
In Fig.~\ref{gate}, we present our results for the RKKY exchange interaction between two magnetic impurities for an undoped ($E_F=0$) ZPNR, as a function of the distance between them. We employ Eq.(\ref{J_equation}) to compute the RKKY exchange interaction. We choose three different impurity configurations within the lattice as mentioned earlier. Here, Fig.~\ref{gate}(a) corresponds to the case when both the impurities are placed at the same edge. We fix one of the impurities at the position $(1,m_1)$ and the location of the second impurity is at $(1,m_2)$. 
In our numerical analysis, we vary $\Delta m$ ($=m_2-m_1$) from $5$ to $40$ (in units of the lattice parameter $b$). We observe that the behavior of $J_{\rm{RKKY}}$ with $\Delta m$ is oscillatory in nature. This oscillatory behavior with distance between the impurities comes out to be very similar to that of other $2$D Dirac materials as reported earlier in the literature~\cite{PhysRevB.81.205416,PhysRevB.94.045443,PhysRevB.96.024413}. The amplitude of the oscillation decays very fast as we increase the distances between the two impurities. However, for the case of nanoribbon where we deal with 
the lattice model instead of continumm as in the bulk, exact functional dependence is difficult to establish. Nevertheless, from our numerical analysis we can only predict
that the pattern of the RKKY interaction exhibits close resemblance to $1/R^3$ decay.
The characteristic feature of RKKY interaction in ZPNR, in absence of gate voltage, is very similar to that of graphene
as discussed in Ref.~[\onlinecite{PhysRevB.81.205416}]. To discuss the effect of external gate potential,
we choose three different values of $U$ ($=0$, $1.5$, and $3$ in units of $t_1$). We observe that 
$J_{\rm{RKKY}}$  attains maximum strength when the applied gate voltage is zero. The RKKY interaction 
becomes vanishingly small with the increase of the gate voltage $U$ 
\begin{figure}[!thpb]
\centering
\includegraphics[height=6.5cm,width=\linewidth]{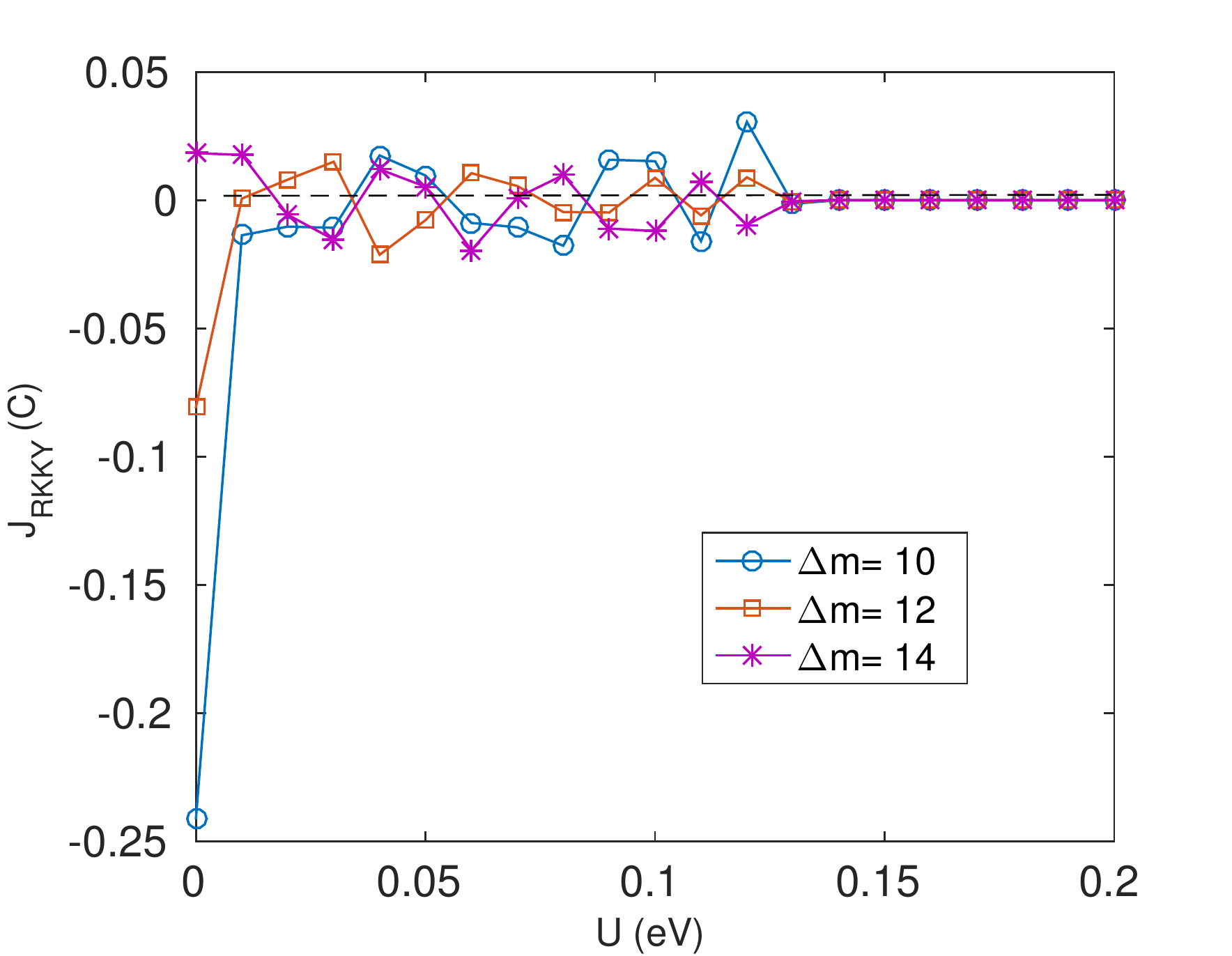}
\caption{(Color online) The feature of RKKY interaction strength (in units of $C$) is illustrated with respect to the gate voltage 
for three different spatial distributions (in units of $b$) of two magnetic impurities.}  
\label{rkky_gate}
\end{figure}
(see Fig.~\ref{gate}(a)). The reason behind this phenomenon can be explained
from the features of band structure as shown in Fig.~\ref{band1}.
\begin{figure*}[!thpb]
\centering
\includegraphics[height=4.5cm,width=0.33\linewidth]{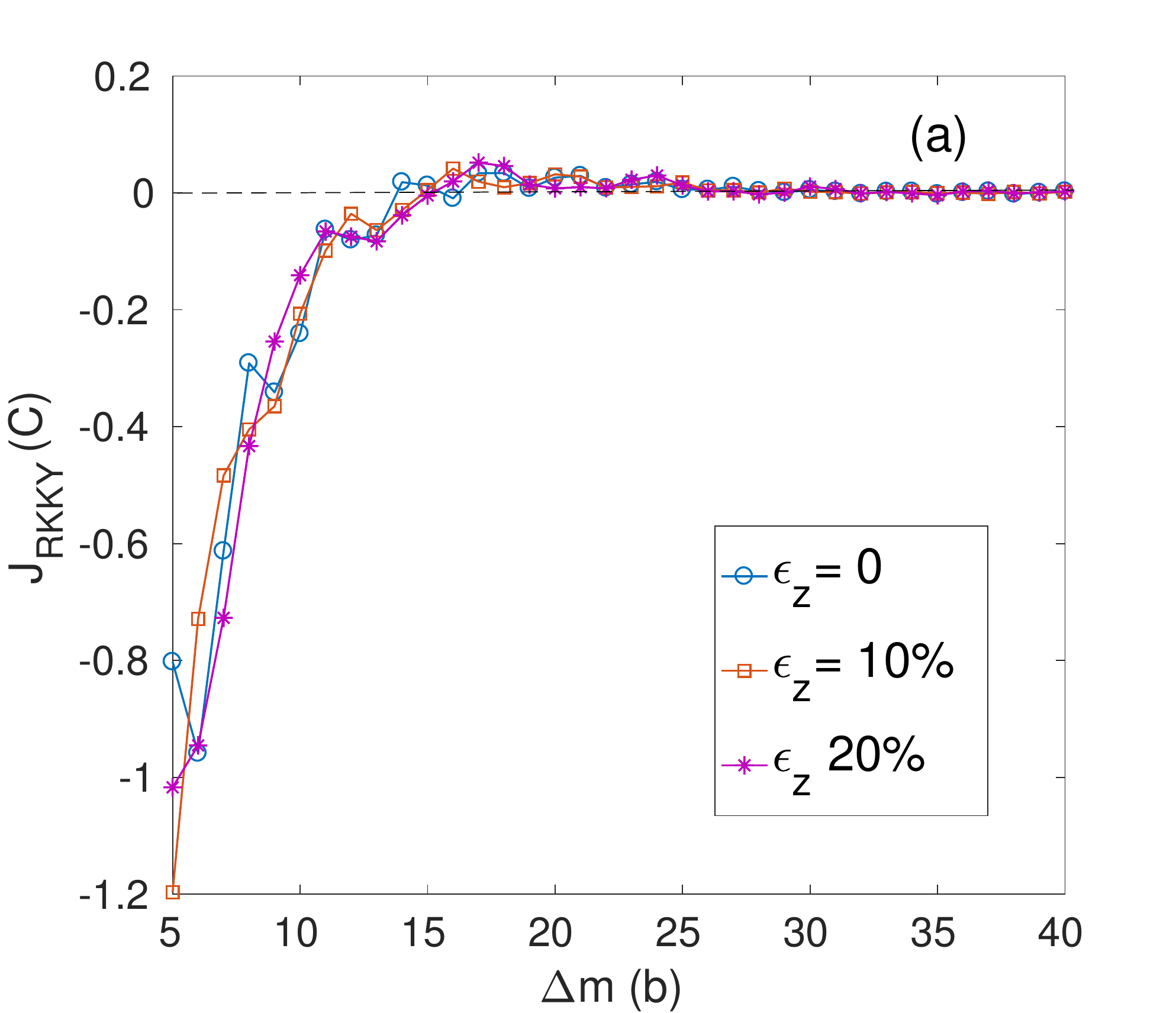}
\includegraphics[height=4.5cm,width=0.32\linewidth]{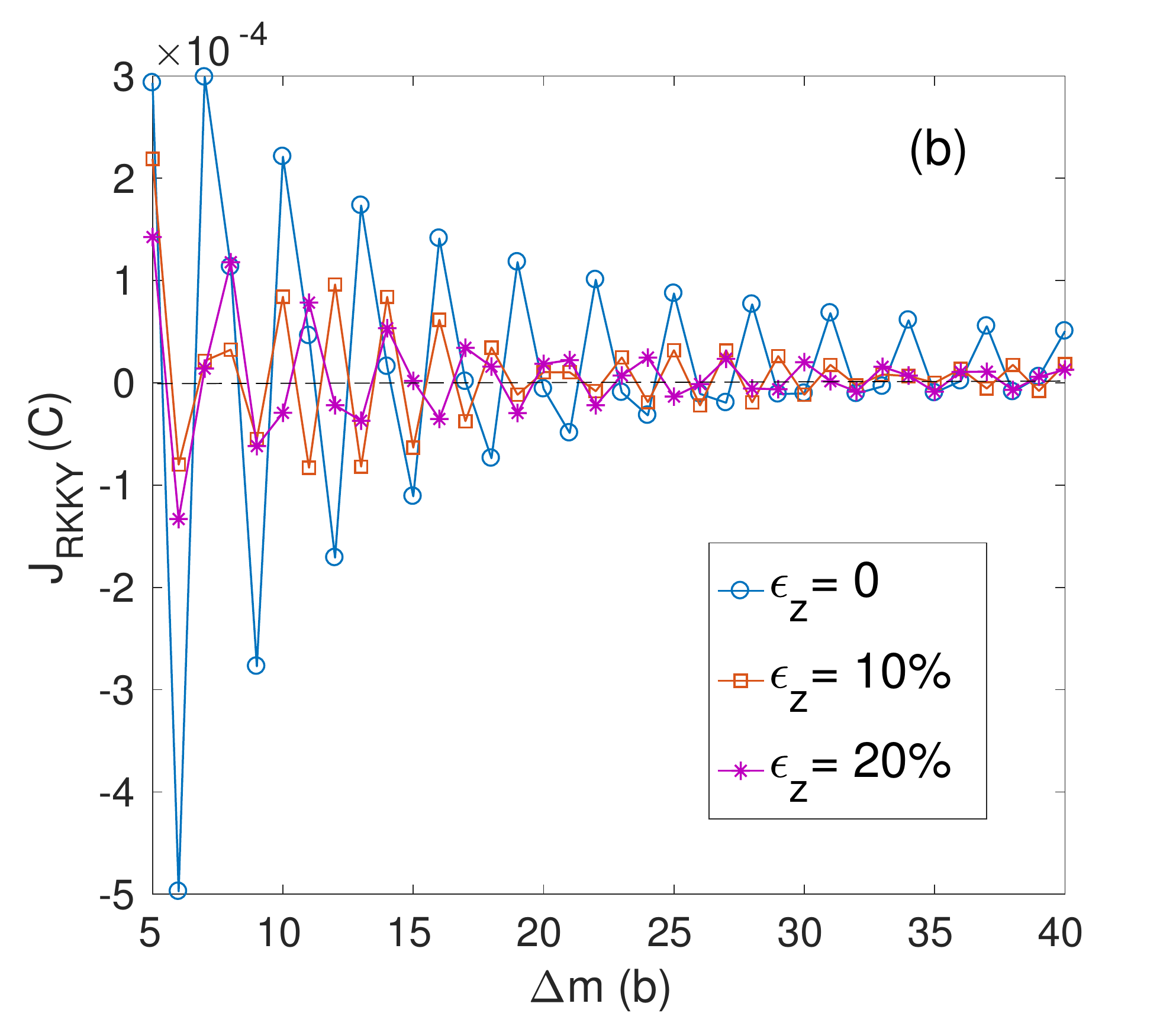}
\includegraphics[height=4.5cm,width=0.33\linewidth]{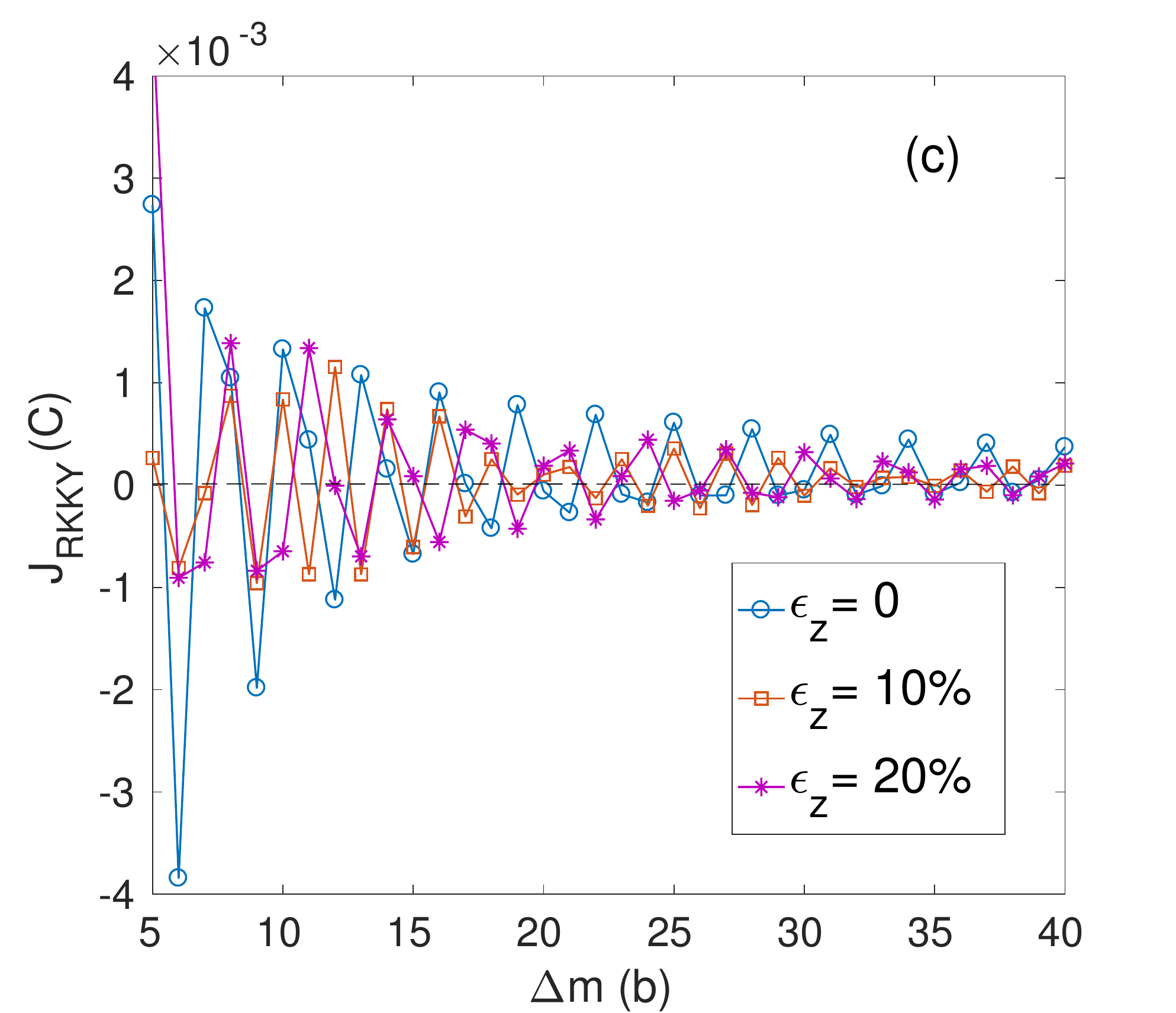}
\caption{(Color online) The behavior of RKKY exchange interaction (in units of $C=(J_c\hbar/2)^2$) between two magnetic impurities is illustrated 
as a function of the distance (in units of lattice parameter $b$) between them when (a) both the impurities are at the same edge, (b) both are within the bulk, 
and (c) one at the edge while the other one within the bulk. The size of the undoped ($E_{F}=0$) ZPNR is $10 \times 100$. For each case, three different values 
of strain, $\eps_z=0$, $10\%$ and $20\%$ (in units of $t_1$) are taken into account in absence of any gate voltage ($U=0$).}
\label{str}
\end{figure*}
For $U=0$, the Fermi level crosses the decoupled edge states. As soon as we switch on the gate voltage 
the Fermi level moves away from the edge modes to the gap between the bulk and edge states where the
density of states are vanishingly small to conduct. As the bands near the Fermi levels contribute to
tunneling exchange, the RKKY interaction strength becomes vanishingly small due to the unavailibility
of the DOS for $U\neq0$. On the other hand, the origin of the RKKY interaction can be purely attributed
to the edge states of the ZPNR when $U=0$. The contribution of the bulk states are almost zero as both
the impurities are located at the edge of the ribbon. Hence, one can separately probe the edge states
of ZPNR via the RKKY interaction. These features of the RKKY interaction can be further analyzed in 
terms of local density of states (LDOS) expressed for $i$-th site as
\begin{equation}
 \rho_i=-\frac{1}{\pi}{\rm Im}[G_{ii}({\bf r,r,}E)] .
\end{equation}
The LDOS is demonstrated in Fig.~\ref{ldos} which manifests the existence of relatively higher LDOS around 
the zero energy corresponding to edge states ($n=1$) in comparison to the bulk ($n=5$). The other peaks in LDOS 
around  $E=+2$ eV and $-2.2$ eV, present in both panels (a) and (b), correspond to the bulk states. The asymmetry around 
zero energy corroborates the particle-hole asymmetry in the band structure described above.  Also, this asymmetry
has been reported earlier in the context of band structure~\cite{PhysRevB.95.045422}.
The central peak is well separated from the bulk for which the RKKY interaction becomes vanishingly small when the Fermi level 
is tuned into the gap between the central and nearest bulk peaks.

Here, we present a comparative analysis between ZPNR and other hexagonal lattices like graphene and silicene
in the context of RKKY interaction. The graphene zigzag nanoribbon also consists of zero energy edge modes
as well as gapless bulk states which are in contrast to ZPNR where the bulk is gapped. Apart from that, the
edge modes are not isolated from the bulk states. Rather they merge into the bulk at the two
valleys~\cite{RevModPhys.81.109}, for which a small deviation from 
the undoped case would not cause any sudden drop in the RKKY interaction as the Fermi level always passes
through the edge modes. Moreover, as the bulk is gapless, the contribution of edge states on the RKKY
interaction in an undoped graphene zigzag nanoribbon will always be accompanied by the bulk states.
In case of silicene~\cite{PhysRevB.92.035413}, although it exhibits gapped bulk band structure due to
the strong spin-orbit interaction owing to buckled lattice structure, the edge modes are not decoupled 
from the bulk. Whereas, the edge modes in ZPNR are fully separated from the bulk which yields a sudden
drop in the RKKY interaction amplitude after a small deviation of the Fermi level from the edge modes
by means of a gate voltage. This unique nature of the edge states in ZPNR allows one to probe them
separately from the bulk. 

In Fig.~\ref{gate}(b), we consider the case when both the impurities are away from the edge. The locations 
of the two magnetic impurities are considered at $(5,m_1)$ and $(5,m_2)$. We vary $\Delta m$ from $5$ to
$40$ (in units of $b$) as mentioned in the previous case. We observe that the RKKY interaction is negligibly
small even the Fermi level passes through the decoupled edge states as shown in Fig.~\ref{band1}. 
As the Fermi level is far away from the bulk states, it leads to vanishingly small contribution to
the exchange interaction in undoped situation when both the impurities are situated in the interior
region of the ZPNR. The DOS due to the edge modes doesn't contribute to RKKY for $U=0$. By the application
of gate voltage, the energy band dispersion inside the bulk as well as the edge modes are shifted and
the bulk states come closer to the Fermi level for which RKKY interaction inside the bulk becomes significant.
At gate voltage $U= 3t_{1}$, the RKKY interaction manifests smooth oscillation with relatively higher amplitude 
(see Fig.~\ref{gate}(b)). Such higher amplitude is the consequence of the availibility of large DOS due to
the bulk bands as the Fermi level crosses them (see Fig.~\ref{band1}). Hence, a clear distinction between 
the nature of the RKKY exchange interactions for bulk and edge modes are now visible. The strength of RKKY
interaction displays a smooth oscillation with higher amplitude when both the impurities are located within
the bulk whereas, it oscillates rapidly and decays very fast in the case when we place them at the edge of the ZPNR.

Finally, we consider the case when one magnetic impurity is situated at the edge and other one is located within the bulk. 
The locations of the two impurities are at $(1,m_1)$ and $(5,m_2)$ and $\Delta m$ is varied as mentioned before. 
The corresponding behavior of RKKY interaction, for this situation, is illustrated in Fig.~\ref{gate}(c).
Here, we observe that the RKKY interaction is dominated by the edge modes when $U=0$. 
However, with the further enhancement of the gate voltage ($U\neq0$), the bulk states also start to contribute 
for which a smooth oscillation with higher frequency appears in the behavior of RKKY.
This oscillation is mostly confined within the regime of positive (ferromagnetic) sign of the interaction 
(see Fig.~\ref{gate}(c) and $U=3t_{1}$ in particular).

Therefore, in all the above three cases, depending on the gate voltage, the interplay of the Fermi level and the 
LDOS (edge/bulk) gives rise to the desirable RKKY exchange interaction between the two magnetic impurities. 
However, the features of exchange interactions still differ from each other in terms of nature of oscillation.

At this stage, we also show how the RKKY interaction behaves with the variation of the gate voltage
in Fig.~\ref{rkky_gate} for fixed distance between the two impurities located at the same edge. 
We observe that when the gate voltage is zero, the RKKY interaction is maximum and it decays very fast associated 
by small fluctuation with the enhancement of gate voltage. Such sharp reduction in amplitude with respect to the gate voltage
is expected as the Fermi level deviates from the decoupled edge states for large $U$. This small contribution with fluctuation 
even for a finite gate voltage is the consequence of small density of states around the edge modes.
Note that, the behavior of the RKKY interaction for different values of $\Delta m (=12,14)$ is almost
similar as far as amplitude and phase mismatch are concerned. This is expected as we have already demonstrated 
previously that the RKKY interaction exhibits rapid fluctuation with the distance between the two magnetic impurities (see Fig.~\ref{gate}(a)).
\subsection{Effect of tensile strain}
In this subsection, we investigate how the strength of RKKY interaction responds to the different
degrees of tensile strain. Considering the same system size, RKKY interaction strength is
numerically computed for various spatial configurations of the two magnetic impurities as mentioned 
in Fig.~\ref{gate}. 

In Fig.~\ref{str}(a), we show the behavior of $J_{\rm{RKKY}}$ as a function of $\Delta m$ for an undoped ZPNR, considering the case 
when both the magnetic impurities are at the same edge. 
We observe that for small separation between the two impurities, the RKKY interaction is strong with only negative sign (anti-ferromagnetic).
However, it decays exponentially fast and becomes vanishingly small as we increase the distance between them.  
Moreover, the RKKY interaction seems to be very weakly sensitive to the degree of strain in this case.
This can be explained by the band structure analysis of the undoped strained ZPNR (see Fig.~\ref{band2}). 
As the Fermi level crosses through the edge states, irrespective of the degree of strain, the amplitude of the RKKY interaction 
remains almost unaffected with the strain. Although the bulk states of the ribbon gets affected by the applied strain, 
it does not reflect in the feature of RKKY interaction as both the impurities are located at the edge of the ribbon. 

Similar to the case of gate voltage, we also consider the situation where both the impurities
are located in the interior of the undoped nanoribbon. Our corresponding results for the RKKY interaction as a function
of $\Delta m$ is depicted in Fig.~\ref{str}(b). The strength of the interaction abruptly falls down in this case
compared to the earlier case where the impurities were situated at the edge. This occurs as the bulk 
states are away from the Fermi level of the undoped ribbon as shown in Fig.~\ref{band2}.
Hence, the available DOS to mediate RKKY interaction in this case is vanishingly small. 
Therefore, although the impurity positions are within the bulk region, the amplitude of the RKKY interaction
is still relatively small as only the edge states being close to the Fermi level can mediate the
exchange interaction. The variation of the degree of strain also does not affect the strength of the 
interaction significantly even the bulk states are deformed substantially with strain. 
The reason is that the bulk states are well separated from the Fermi level by a substantial gap even with $\eps_z=20\%$.
However, the enhancement of strain induces a small phase to the oscillation of the RKKY interaction. 
Note that, as far as the oscillatory nature of the interaction in concerned, switching the phase from ferromagnetic 
to the anti-ferromagnetic order and vice versa is still present. This feature is very similar to the other
$2$D materials~\cite{PhysRevB.81.205416,PhysRevB.94.045443}.
In Fig.~\ref{str}(c), we demonstrate the behavior of RKKY interaction for the case when one impurity is at the edge and the 
other one is inside the bulk of the undoped ZPNR. This also manifests oscillatory behavior with distance between the two impurities. 
Such oscillatory nature as well as the amplitude of the interaction are almost insensitive to the strain. However, the strength 
of the interaction increases in comparison to the case shown in Fig.~\ref{str}(b). This happens as one of the
impurities are located at the edge of the ribbon and the available DOS due to the edge states contributes to 
the finite value of $J_{\rm{RKKY}}$.

\begin{figure}[!thpb]
\centering
\includegraphics[height=6.0cm,width=\linewidth]{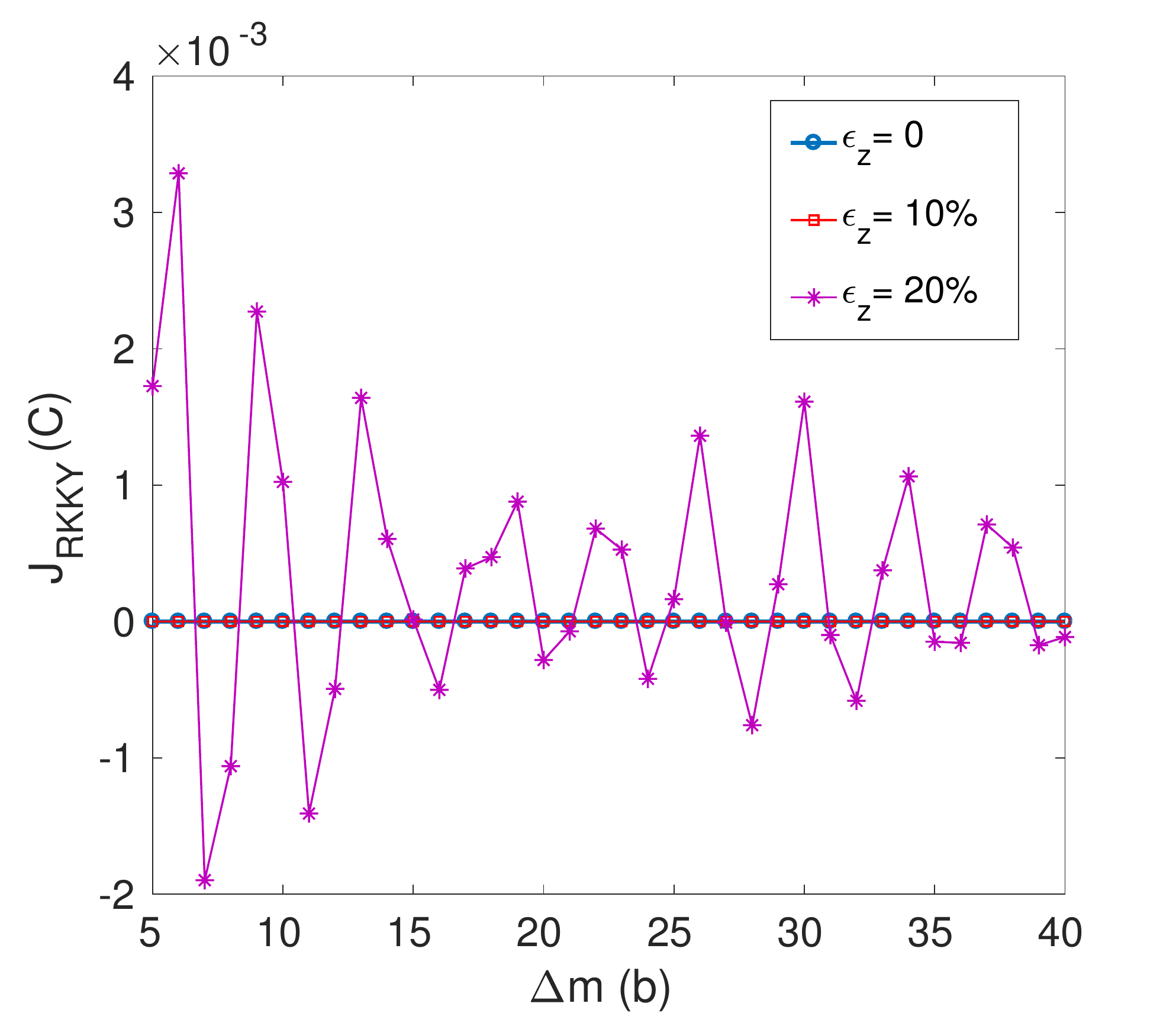}
\caption{(Color online) The behavior of RKKY interaction strength (in units of $C$) between two impurities is illustrated as a function of 
distance (in units of $b$) between them when both the impurities are inside the bulk. The ZPNR is doped at $E_F=1.6 t_1$
and $U=0$. We choose the same size of the ribbon as mentioned in Fig.~\ref{str}.}
\label{str_doped}
\end{figure}
So far, we have considered undoped ribbon ($E_F=0$). From the band structure shown in Fig.~\ref{band2},
we can conclude that the RKKY interaction strength may enhance significantly if one dopes the system locating
both the impurities within the bulk region of the ZPNR. We illustrate the behavior of $J_{\rm RKKY}$ as a function
of the relative separation between the impurities in Fig.~\ref{str_doped}. We choose three different values of the
strain ($0$, $10\%$ and $20\%$), after tuning the Fermi level at $E_F=1.6$ (in units of $t_1$). We observe that
the RKKY interaction strength in the doped ZPNR increases significantly when we apply high degree of strain. 
For example, the exchange interaction becomes very strong for the strain of $20\%$ in comparison to zero and $10\%$.
The reason can be attributed to the band dispersion (see Fig.~\ref{band2}) which exhibits that Fermi level lies far away
from the bulk states when strain is considered to be at zero and $10\%$. However, as we apply strong degree of 
strain ($20\%$) then it induces a strong curvature to the bulk bands and effectively reduces the band gap between 
the bulk states. Thus, the Fermi level intersects the bulk bands. This induces a sizable contribution to the RKKY
interaction between the two magnetic impurities positioned inside the bulk region of the dopped ZPNR. 
Note that, RKKY exchange interaction also exhibits a beating pattern around $\Delta m=20$.  
This appears due to the superposition of two contributions arising from the two closely spaced different momenta, 
as the Fermi level passes through them (see Fig.~\ref{band2}(c) for illustration).
\subsection{Combined effect of gate voltage and strain}
\begin{figure}[!thpb]
\centering
\includegraphics[height=6.0cm,width=\linewidth]{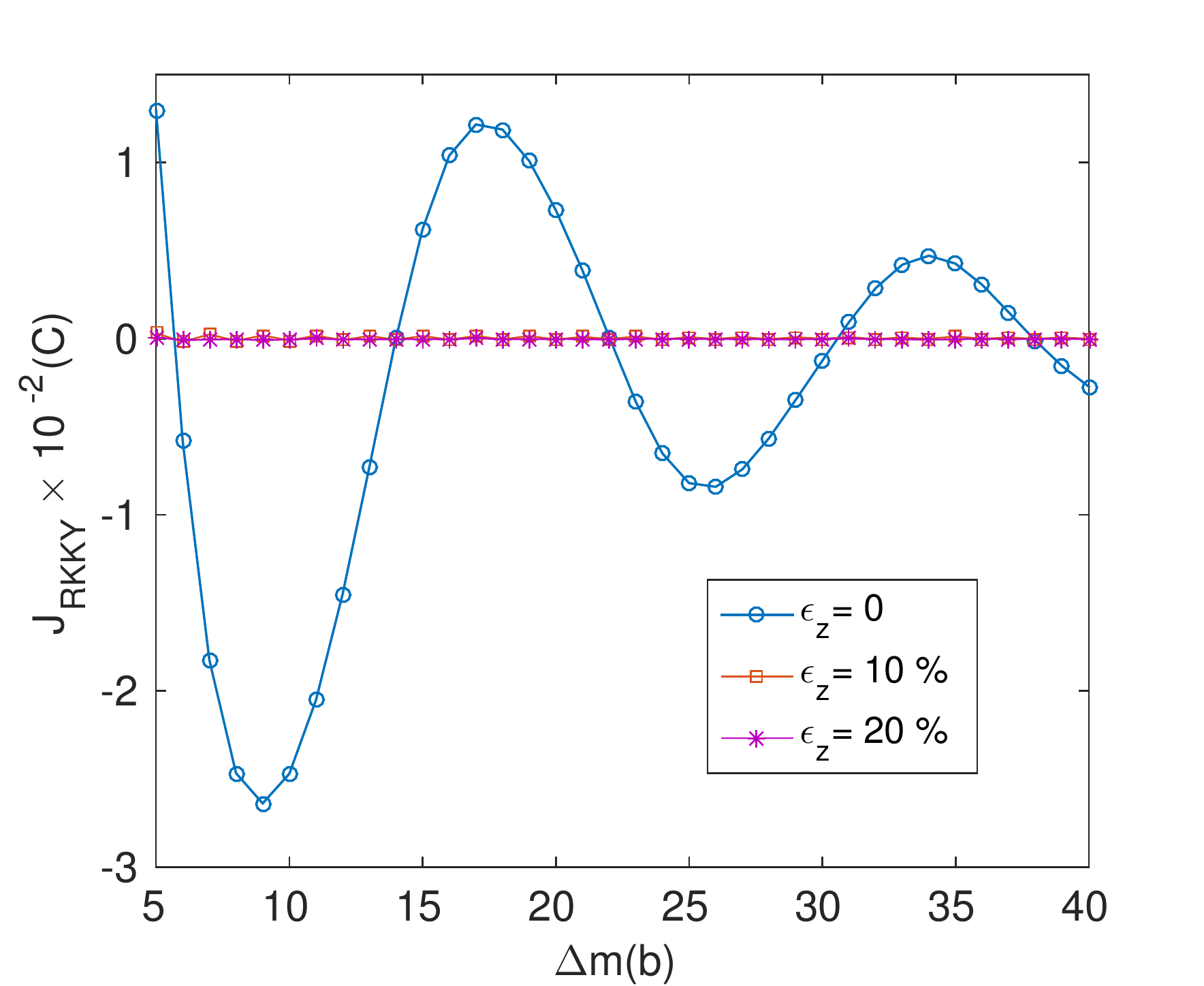}
\caption{(Color online) The feature of RKKY exchange interaction (in units of $C$) between two magnetic impurities, 
located inside the bulk of a ZPNR, is demonstrated as a function of the distance (in units of $b$) between them. 
We choose $U=3t_1$, $E_F=0$ and same size of the ribbon as mentioned in Fig.~\ref{str}.}
\label{gate_str_both_bulk}
\end{figure}

Here, we consider the case when both the gate voltage and the strain are applied together to the undoped ($E_{F}=0$) ZPNR.
We present our results of the RKKY exchange interaction as a function of the spatial separation between the impurities in 
Fig.~\ref{gate_str_both_bulk}, for three different values of the strain. We also consider non-zero gate voltage at
$U=3t_1$ for which the Fermi level lies very close to the bulk band and far away from the edge modes (see Fig.~\ref{band3}). 
Therefore, this configuration gives rise to the dominant contribution in the RKKY exchange only when both the
impurities are inside the bulk. Also, the exchange interaction becomes vanishingly small when one of the two
impurities or both the impurities reside at the same edge due to the unavailibility of sufficient DOS to mediate
RKKY.
Furthermore, in this case, we observe a smooth oscillation in the behavior of $J_{\rm{RKKY}}$ for
$\epsilon_{z}=0$ which is already discussed in the earlier subsection in the context of gate voltage 
(see Fig.~\ref{gate}(b)). However, with the enhancement of strain, the RKKY interaction suddenly drops to zero.
The reason can be attributed to the fact that as we increase the degree of strain, the band gap between the 
conduction and valence band reduces, and subsequently the Fermi level is repositioned much inside the bulk states.
The strength of the RKKY interaction is inversely proportional to the Fermi momentum. This causes sudden drops
in exchange interaction strength when we tune the strain to $10\%$ or $20\%$.
\begin{figure}[!thpb]
\centering
\includegraphics[height=6.5cm,width=\linewidth]{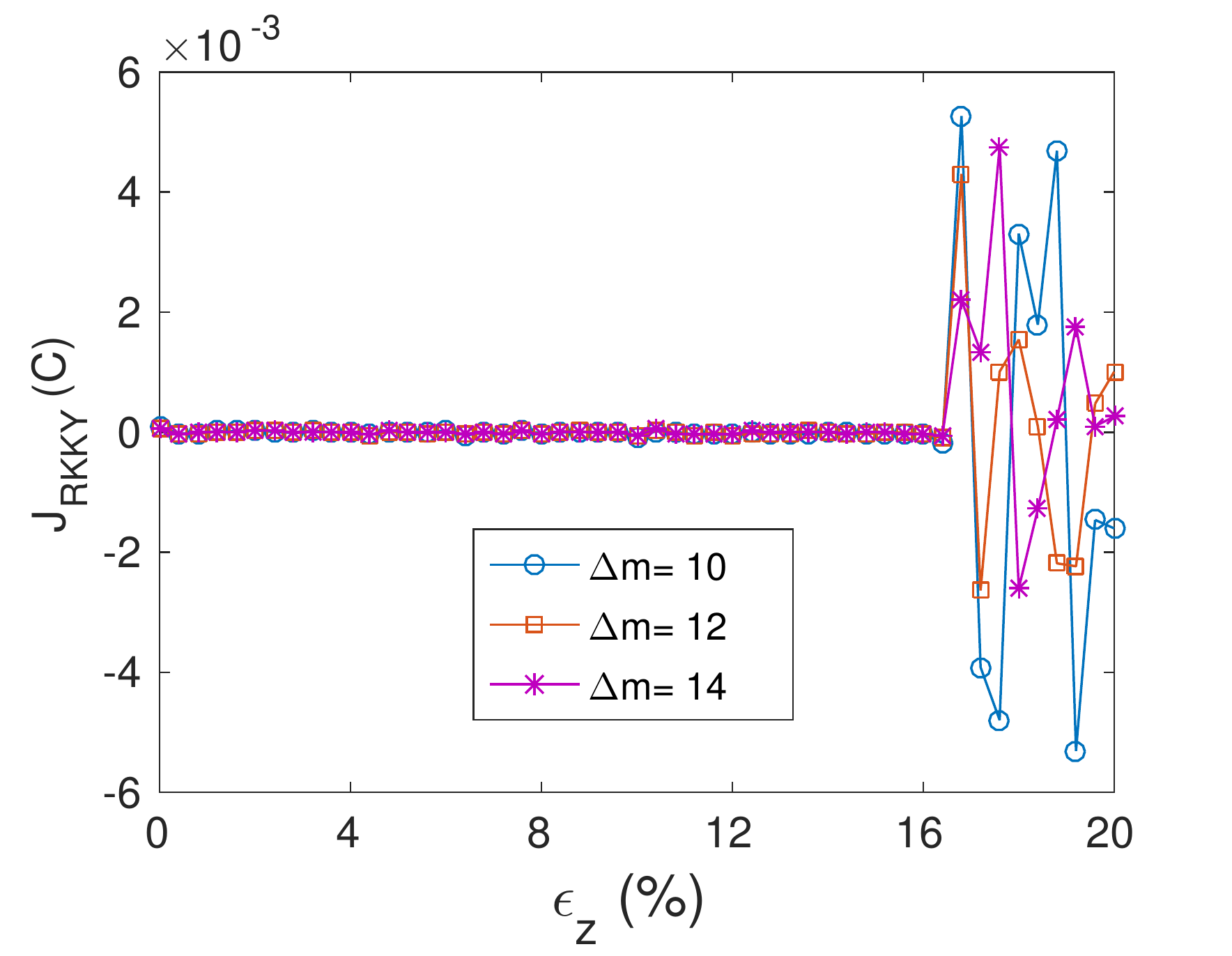}
\caption{(Color online) The behavior of RKKY interaction strength (in units of $C$) is illustrated as a function of degree of strain 
for three different spatial distributions (in units of $b$) of two  impurities. We consider finite doping $E_{F}=-2$ eV and 
gate voltage $U=3$ eV.}  
\label{rkky_strain}
\end{figure}
The corresponding band dispersion in presence of both gate voltage and strain, depicted in Fig.~\ref{band3}, manifests that Fermi level  $E_{F}$ intersects the bulk bands with higher momentum for higher degree of strain and consequently it weakens the exchange interaction mediated through the conduction electrons.

Similar to the plot of RKKY interaction with respect to gate voltage shown in Fig.~\ref{rkky_gate}, we here depict the behavior of RKKY interaction strength 
with the variation of strain in Fig.~\ref{rkky_strain}. We choose three different values of $\Delta m$ and a particular value of gate voltage $U$. 
To obtain better signatures of strain on the RKKY interaction we have chosen ZPNR with finite doping. The corresponding band dispersion for that case is shown in Fig.~\ref{band3}.
The RKKY interaction remains vanishingly small till the degree of strain reaches $\epsilon_z=16$\% when the edge modes start overlapping with the Fermi level. 
Beyond this degree of strain, the RKKY interaction shows sudden rise with large fluctuations. A general statement regarding such behavior of the amplitude of the RKKY interaction 
can be attributed to the interplay of Fermi level with the band dispersion (edge modes or bulk states). 

\section{Summary and conclusions}\label{sec6}
To summarize, in this article, we numerically investigate the RKKY exchange interaction between two magnetic impurities
located on a zigzag phosphorene nanoribbon. The signatures of quasi-flat edge modes, via RKKY interaction, in ZPNR have 
been explored. We show that the small deviation of the Fermi level, by means of gate voltage, gives rise to
a sudden drop in the strength of the RKKY interaction between the two magnetic impurities positioned at the same 
edge. Note that, this sudden drop is a consequence of the separation between the edge and the bulk states and LDOS therein.
When the Fermi level lies within the gap between the edge and bulk states, the density of states of the conduction 
electrons is negligibly small to contribute significantly to the RKKY exchange phenomenon. On the other hand, in other
$2$D Dirac materials like graphene and silicene zigzag nanoribbon, we cannot separately identify the contributions of 
the edge states as they merge inside the bulk bands. In undoped graphene, the contribution of edge modes to the 
RKKY interaction is always accompanied by the bulk contribution \ie they are inseparable. Whereas in silicene, although
it possesses a band gap due to the spin-orbit interaction, the edge modes are not fully decoupled from the bulk and
merge inside the bulk bands at two valleys. Hence, one cannot separate out their contributions to the RKKY interaction too. 
Therefore, phosphorene is a semi-Dirac material in which the separation of quasi-flat edge modes and the isolation of its 
contribution via the RKKY exchange interaction can be a possible probe to detect them in a ribbon geometry. 
Moreover, the nature of the oscillations in the RKKY interaction are in complete contrast to each other in ZPNR, 
when both the impurities are in the bulk or at the same edge. 
We also consider the effect of tensile strain on RKKY exchange interaction. The strain does not impart any shift to the band dispersion, 
rather it attributes a curvature to the bulk bands. Such curvature introduces an additional phase shift to the RKKY oscillation 
with the distance between the two impurities. The amplitude of the exchange interation is weakly sensitive to the strain value. 
However, one can enhance the strength of the interaction by adjusting the Fermi level at suitable position. Finally, we also explore
the case when both gate voltage and strain are applied simultaneously to the ZPNR. In this case, the amplitude as well as the 
oscillation of the interaction profile is very sensitive to the Fermi energy too.

\begin{acknowledgments}
PD thanks Department of Science and Technology (DST), India for the financial support through SERB NPDF (File no. PDF/2016/001178). 
AMJ also thanks DST, India for financial support through JC Bose fellowship.
\end{acknowledgments}
\bibliography{bibfile}{}
\end{document}